\documentclass[aps,pra,floatfix,showpacs,twocolumn,tightenlines]{revtex4}
\newcommand{\be}{\begin{equation}}
\newcommand{\ee}{\end{equation}}
\newcommand{\brr}{\begin{eqnarray}}
\newcommand{\err}{\end{eqnarray}}
\newcommand{\nn}{\nonumber}
\newcommand{\bd}{\begin{displaymath}}
\newcommand{\ed}{\end{displaymath}}

\newcommand{\bib}{\bibitem}
\newcommand{\bfig}{\begin{figure}}
\newcommand{\efig}{\end{figure}}
\newcommand{\ie}{i.e.}
\newcommand{\eg}{e.g.}

\newtheorem{corollary}{Corollary}[section]
\newtheorem{lemma}{Lemma}[section]

\newtheorem{proof}{Proof}[section]
\def\alf{\alpha}
\def\bet{\beta}
\def\gam{\gamma}
\def\th{\theta}

\def\om{\omega}
\def\eps{\epsilon}
\def\rpar{\right)}
\def\lpar{\left(}
\def\rbk{\right]}
\def\lbk{\left[}

\def\lb{\label}
\def\ro{\mbox{\boldmath $\rho$}}

\def\bgam{\mbox{\boldmath $\Gamma$}}
\def\hei{\mbox{\tiny ${\rm H}$}}
\def\epr{\mbox{\tiny ${\rm EPR}$}}
\def\inda{\mbox{\tiny ${\rm A}$}}
\def\indb{\mbox{\tiny ${\rm B}$}}
\def\rg{\rangle}
\def\lgg{\langle}
\def\nc{\mathrm{i}}
\def\coloneq{\mathrel{\mathop:}=}
\def\half{\frac{1}{2}}

\usepackage{amssymb}
\usepackage{mathrsfs}
\usepackage{upmath}
\usepackage{yfonts}
\usepackage{dsfont}
\usepackage{amsmath}
\usepackage{graphicx}
\usepackage{bm}
\usepackage{amsfonts}
\usepackage{subfigure}
\usepackage{eucal}
\begin{document}
\title{Generalized squeezing operators, bipartite Wigner functions, and entanglement via Wehrl's entropy functionals}
\author{Marcelo A. Marchiolli}\email{mamarchi@ift.unesp.br}
\author{Di\'{o}genes Galetti}\email{galetti@ift.unesp.br}
\affiliation{Instituto de F\'{\i}sica Te\'{o}rica, Universidade Estadual Paulista, Rua Pamplona 145, 01405-900, S\~{a}o Paulo, SP, Brazil}
\date{\today}
\begin{abstract}
We introduce a new class of unitary transformations based on the $\mathfrak{su}(1,1)$ Lie algebra that generalizes, for certain 
particular representations of its generators, well-known squeezing transformations in quantum optics. To illustrate our results, we
focus on the two-mode bosonic representation and show how the parametric amplifier model can be modified in order to generate such a
generalized squeezing operator. Furthermore, we obtain a general expression for the bipartite Wigner function which allows us to 
identify two distinct sources of entanglement, here labelled by dynamical and kinematical entanglement. We also establish a quantitative
estimate of entanglement for bipartite systems through some basic definitions of entropy functionals in continuous phase-space 
representations.
\end{abstract}
\pacs{02.20.Qs, 03.65.Ca, 03.67.Mn} 
\maketitle
\section{Introduction}

In the last few years, physics has experienced the appearance of two relatively young branches with strong appeal in both theoretical
and experimental aspects. Labelled by Quantum Information and Quantum Computation, these branches apparently entangled have attracted
since then a lot of attention from researchers working in well-established areas in physics (such as, for instance, solid state physics,
nuclear physics, high energy physics, general relativity, and cosmology \cite{Galindo}). The interdisciplinarity provided by Quantum 
Information and Quantum Computation is basically focussed upon fundamental physical concepts that constitute the cornerstones of quantum 
mechanics. Hence, concepts related to nonclassical states, superposition principle, entanglement, phase-space representations, quantum 
teleportation, quantum key distribution, among others, represent nowadays common words in many scientific papers covering different areas 
of knowledge in physics. 

In particular, let us restrict our attention to quantum information theory and its description by quantum continuous variables 
\cite{Braunstein}, where entanglement effects can be efficiently produced in laboratory through the adequate manipulation of continuous 
quadrature amplitudes of the quantized electromagnetic field. In this promising scenario, the squeezed light \cite{Books,Leonhardt} has a 
prominent role in the experimental implementation of continuous-variable entanglement, since the degree of imperfection in entanglement-based 
quantum protocols depends on `the amount of squeezing of the laser light involved'. From a conceptual point of view, some questions related 
to the entanglement measures (or separability criteria) still remain open in the specialized literature \cite{Horodecki}. For instance, 
entanglement of formation \cite{Bennett} and concurrence \cite{Wootters} are now widely accepted as entanglement measures for the two-qubit 
case. However, it is worth noticing that different approaches to this problem exist which are outlined by means of information-theoretic 
arguments \cite{Adami}.

The main goal of this paper is to present some contributions to certain specific topics in quantum information theory that allow us to
go further in our comprehension on the entanglement process in ideal bipartite systems (the interaction with any dissipative environment
\cite{Breuer} is discarded in a first moment). For this purpose, we first construct a general family of unitary transformations
associated with the $\mathfrak{su}(1,1)$ Lie algebra generators that generalizes -- if one considers the one- and two-mode bosonic
representations of its generators -- two well-known expressions of squeezing operators \cite{Books}. Following, we study a modified
version of the parametric amplifier \cite{Yariv} with emphasis in obtaining the solutions of the Heisenberg equations and its respective
time-evolution operator. In particular, we show the efficacy of this model in generating the generalized two-mode squeezing operator
through its connection with the time-evolution operator. 

The next step then consists in performing a preliminary study via Wigner function on the qualitative aspects of entanglement for a general 
class of bipartite systems whose dynamics is governed by arbitrary quadratic Hamiltonians. As a by-product of this investigation, we obtain 
a general integral representation for the bipartite Wigner function that leads us to identify two distinct sources of entanglement (here 
labelled by dynamical and kinematical entanglement). The last contribution refers to a direct application of the results obtained by 
Pi\c{a}tek and Leo\'{n}ski \cite{Piatek} for the intermode correlations in continuous phase-space representations, which are based on the 
Wehrl's approach \cite{Wehrl} regarding some definitions of entropy functionals and their inherent properties. In fact, we introduce a 
correlation functional that permits us to measure the degree of entanglement between both parts of the joint system for any initial 
conditions associated with the Hamiltonian operator. It is important to emphasize that the sequence of topics covered in this work presents 
an inherent logical consistency that improves our comprehension on the subtle mechanisms associated with the entanglement effects in 
bipartite systems.

This paper is structured as follows. In Section II, we present a mathematical statement that leads us to construct a general family of
unitary transformations associated with the $\mathfrak{su}(1,1)$ Lie algebra where, in particular, the two-mode bosonic representation
is emphasized. In Section III, we obtain the solutions of the Heisenberg equations for the parametric amplifier model, and show the subtle
link between time-evolution operator and generalized two-mode squeezing operator. The results obtained are then applied in Section IV, 
within the context of Wigner functions, in order to establish an initial discussion on entanglement for certain groups of bipartite 
states of the electromagnetic field. Section V is devoted to establish a reasonable measure of entanglement which is based on some specific 
information-theoretic arguments. Besides, we illustrate our results through two different examples of initially uncoupled bipartite states 
for the model under investigation. Finally, Section VI contains our summary and conclusions. 

\section{Squeezing operators associated with the $\mathfrak{su}(1,1)$ Lie algebra}

Let us initially introduce the generators of the $\mathfrak{su}(1,1)$ Lie algebra $\mathbf{K}_{\pm}$ and $\mathbf{K}_{0}$, which
satisfy the following commutation relations: 
\bd
\lbk \mathbf{K}_{-},\mathbf{K}_{+} \rbk = 2 \mathbf{K}_{0} \quad \mbox{and} \quad \lbk \mathbf{K}_{0},\mathbf{K}_{\pm} \rbk = \pm 
\mathbf{K}_{\pm} \; . 
\ed
The Casimir operator is defined within this context through the mathematical identity
\bd
\mathbf{K}^{2} \coloneq \mathbf{K}_{0}^{2} - (1/2) \{ \mathbf{K}_{-},\mathbf{K}_{+} \} 
\ed
with $[ \mathbf{K}^{2},\mathbf{K}_{\pm} ] = [ \mathbf{K}^{2},\mathbf{K}_{0} ] = 0$, where $\{ \mathbf{K}_{-},\mathbf{K}_{+} \}$ represents 
the anticommutation relation between the operators $\mathbf{K}_{-}$ and $\mathbf{K}_{+}$. Furthermore, let us also consider the abstract 
operator
\be
\lb{e1}
{\bf T}(\Omega_{\pm},\Omega_{0}) \coloneq e^{\Omega_{+} \mathbf{K}_{+} + \Omega_{0} \mathbf{K}_{0} + \Omega_{-} \mathbf{K}_{-}}
\ee
written in terms of the arbitrary c-number parameters $\Omega_{\pm}$ and $\Omega_{0}$, whose generalized normal- and antinormal-order
decomposition formulas have already been established in literature \cite{Eberly,Ban}. These preliminary considerations on the 
$\mathfrak{su}(1,1)$ Lie algebra lead us to demonstrate an important composition formula involving the product of two abstract operators, 
each one being defined as in equation (\ref{e1}) and characterized by a particular set of arbitrary c-number parameters. In fact, this 
section will deal with some applications of the composition formula with emphasis on unitary transformations which resemble, for certain 
representations of the generators ${\bf K}_{\pm}$ and ${\bf K}_{0}$, the squeezing transformations in quantum optics.
\begin{lemma}
Let ${\bf T}(\Omega_{\pm},\Omega_{0})$ and ${\bf T}(\Lambda_{\pm},\Lambda_{0})$ be two abstract operators whose functional forms obey
equation (\ref{e1}). For a given set of arbitrary c-number parameters $\{ \Omega_{\pm},\Omega_{0},\Lambda_{\pm},\Lambda_{0} \}$ it is
always possible to verify the general composition formula
\be
\lb{e2}
{\bf T}(\Omega_{\pm},\Omega_{0}) {\bf T}(\Lambda_{\pm},\Lambda_{0}) = {\bf T}(\Sigma_{\pm},\Sigma_{0}) ,
\ee
where $\Sigma_{\pm}$ and $\Sigma_{0}$ are solutions of the coupled set of nonlinear equations
\begin{subequations}
\begin{align}
\lb{e3a}
\frac{A_{+} + B_{0} B_{+} \lbk A_{0} - A_{+} \lpar A_{-} + B_{-} \rpar \rbk}{B_{0} \lpar A_{-} + B_{-} \rpar} = 
\frac{\Sigma_{+}}{\Sigma_{-}} , \\
\lb{e3b}
\sqrt{B_{0} / A_{0}} \lpar A_{-} + B_{-} \rpar = \lpar \Sigma_{-} / \bet \rpar \sinh (\bet) , 
\end{align}
\end{subequations}
with $\bet = [ (\Sigma_{0}/2)^{2} - \Sigma_{+} \Sigma_{-} ]^{1/2}$. The c-number functions $A_{\pm}$, $A_{0}$, $B_{\pm}$, and $B_{0}$, 
present in the lhs of equations (\ref{e3a}) and (\ref{e3b}), are connected with $\{ \Omega_{\pm},\Omega_{0},\Lambda_{\pm},\Lambda_{0} \}$ 
through the identities \cite{Ban}:
\begin{subequations}
\begin{align}
\lb{e4a}
A_{\pm} &= \frac{\lpar \Omega_{\pm} / \phi \rpar \sinh (\phi)}{\cosh (\phi) - \lpar \Omega_{0} / 2 \phi \rpar \sinh (\phi)}, \\
\lb{e4b}
A_{0} &= \lbk \cosh (\phi) - \lpar \Omega_{0} / 2 \phi \rpar \sinh (\phi) \rbk^{-2}, \\
\lb{e4c}
B_{\pm} &= \frac{\lpar \Lambda_{\pm} / \th \rpar \sinh (\th)}{\cosh (\th) + \lpar \Lambda_{0} / 2 \th \rpar \sinh (\th)}, \\
\lb{e4d}
B_{0} &= \lbk \cosh (\th) + \lpar \Lambda_{0}/2 \th \rpar \sinh (\th) \rbk^{2},
\end{align}
\end{subequations}
for 
\brr
\phi &=& [ (\Omega_{0}/2)^{2} - \Omega_{+} \Omega_{-} ]^{1/2} , \nn \\
\th &=& [ (\Lambda_{0}/2)^{2} - \Lambda_{+} \Lambda_{-} ]^{1/2} . \nn
\err
\end{lemma}
\begin{proof}
Firstly, we apply the generalized normal- and antinormal-order decomposition formulas established by Ban \cite{Ban} for exponential 
functions of the generators of $\mathfrak{su}(1,1)$ Lie algebra to the abstract operators ${\bf T}(\Omega_{\pm},\Omega_{0})$ and 
${\bf T}(\Lambda_{\pm},\Lambda_{0})$, that is,
\be
\lb{e5}
{\bf T}(\Omega_{\pm},\Omega_{0}) = e^{A_{+} {\bf K}_{+}} e^{\ln \lpar A_{0} \rpar {\bf K}_{0}} e^{A_{-} {\bf K}_{-}} 
\ee
and
\be
\lb{e6}
{\bf T}(\Lambda_{\pm},\Lambda_{0}) = e^{B_{-} {\bf K}_{-}} e^{\ln \lpar B_{0} \rpar {\bf K}_{0}} e^{B_{+} {\bf K}_{+}} ,
\ee
where the c-number functions $A_{\pm}$, $B_{\pm}$, $A_{0}$, and $B_{0}$ were defined by means of the identities (\ref{e4a})-(\ref{e4d}). 
In this way, the product ${\bf T}(\Omega_{\pm},\Omega_{0}) {\bf T}(\Lambda_{\pm},\Lambda_{0})$ can be expressed as
\brr
\lb{e7}
{\bf T}(\Omega_{\pm},\Omega_{0}) {\bf T}(\Lambda_{\pm},\Lambda_{0}) &=& e^{A_{+} {\bf K}_{+}} e^{\ln \lpar A_{0} \rpar {\bf K}_{0}} 
e^{\lpar A_{-} + B_{-} \rpar {\bf K}_{-}} \nn \\
& & \times \, e^{\ln \lpar B_{0} \rpar {\bf K}_{0}} e^{B_{+} {\bf K}_{+}} \nn \\
&=& e^{A_{+} {\bf K}_{+}} e^{\ln \lpar A_{0} \rpar {\bf K}_{0}} e^{C_{+} {\bf K}_{+}} \nn \\
& & \times \, e^{\ln \lpar C_{0} \rpar {\bf K}_{0}} e^{C_{-} {\bf K}_{-}} , 
\err
with $C_{\pm}$ and $C_{0}$ given by \cite{Ban}:
\brr
C_{+} &=& \frac{B_{0} B_{+}}{1 - B_{0} B_{+} \lpar A_{-} + B_{-} \rpar} , \nn \\
C_{-} &=& \frac{B_{0} \lpar A_{-} + B_{-} \rpar}{1 - B_{0} B_{+} \lpar A_{-} + B_{-} \rpar} , \nn \\
C_{0} &=& \frac{B_{0}}{\lbk 1 - B_{0} B_{+} \lpar A_{-} + B_{-} \rpar \rbk^{2}} . \nn 
\err
The second step consists in using the relationship
\bd
e^{\ln \lpar A_{0} \rpar {\bf K}_{0}} e^{C_{+} {\bf K}_{+}} = e^{A_{0} C_{+} {\bf K}_{+}} e^{\ln \lpar A_{0} \rpar {\bf K}_{0}}
\ed
for the second and third exponentials in the second equality on the rhs of equation (\ref{e7}), with the aim of establishing the 
intermediate result
\bd
{\bf T}(\Omega_{\pm},\Omega_{0}) {\bf T}(\Lambda_{\pm},\Lambda_{0}) = e^{\lpar A_{+} + A_{0} C_{+} \rpar {\bf K}_{+}}
e^{\ln \lpar A_{0} C_{0} \rpar {\bf K}_{0}} e^{C_{-} {\bf K}_{-}} .
\ed
The rhs of this equation represents the normal-order decomposition of an abstract operator defined as 
\bd
{\bf T}(\Sigma_{\pm},\Sigma_{0}) \coloneq e^{\Sigma_{+} {\bf K}_{+} + \Sigma_{0} {\bf K}_{0} + \Sigma_{-} {\bf K}_{-}} , 
\ed
where the c-number parameters $\Sigma_{\pm}$ and $\Sigma_{0}$ satisfy the following mathematical relations:
\brr
A_{+} + A_{0} C_{+} &=& \frac{\lpar \Sigma_{+} / \bet \rpar \sinh(\bet)}{\cosh (\bet) - \lpar \Sigma_{0} / 2 \bet \rpar 
\sinh (\bet)} , \nn \\
C_{-} &=& \frac{\lpar \Sigma_{-} / \bet \rpar \sinh(\bet)}{\cosh (\bet) - \lpar \Sigma_{0} / 2 \bet \rpar \sinh (\bet)} , \nn \\
A_{0} C_{0} &=& \lbk \cosh (\bet) - \lpar \Sigma_{0} / 2 \bet \rpar \sinh (\bet) \rbk^{-2} , \nn
\err
with $\bet = [ (\Sigma_{0}/2)^{2} - \Sigma_{+} \Sigma_{-} ]^{1/2}$. Consequently, substituting the definitions of $C_{\pm}$ and $C_{0}$ 
in these relations, we obtain a coupled set of nonlinear equations that permits us not only to establish a link between $\{ \Sigma_{\pm},
\Sigma_{0} \}$ and $\{ \Omega_{\pm},\Omega_{0},\Lambda_{\pm},\Lambda_{0} \}$, but also to verify the general composition formula 
(\ref{e2}). $\blacksquare$
\end{proof}

An interesting consequence from this mathematical statement is associated with the construction process of a general family of unitary
transformations where the abstract operator (\ref{e1}) has a central role. To carry out this task let us first establish a corollary
which is directly related to the Lemma II.1.

\begin{corollary}
For $\Omega_{+} = \xi$, $\Omega_{-} = -\xi^{\ast}$, and $\Omega_{0} = \nc \om$, with $\xi \in \mathbb{C}$ and $\om \in \mathbb{R}$, the
abstract operator (\ref{e1}) represents a generator of unitary transformations associated with the $\mathfrak{su}(1,1)$ Lie algebra.
\end{corollary}

\begin{proof}
Basically, the idea is obtaining a specific subset of arbitrary c-number parameters such that 
\bd
{\bf T}(\Omega_{\pm},\Omega_{0}) \lbk {\bf T}(\Omega_{\pm},\Omega_{0}) \rbk^{\dag} = \lbk {\bf T}(\Omega_{\pm},\Omega_{0}) \rbk^{\dag} 
{\bf T}(\Omega_{\pm},\Omega_{0}) = {\bf 1} . 
\ed
Hence, let us initially investigate under what circumstances the mathematical relation 
\bd
{\bf T}(\Omega_{\pm},\Omega_{0}) \lbk {\bf T}(\Omega_{\pm},\Omega_{0}) \rbk^{\dag} = {\bf 1}
\ed
is verified. In fact, this condition can be promptly derived from the general composition formula (\ref{e2}) for $\Lambda_{\pm} = 
\Omega_{\mp}^{\ast}$, $\Lambda_{0} = \Omega_{0}^{\ast}$, and $\Sigma_{\pm} = \Sigma_{0} = 0$, with the additional restrictions 
$\Lambda_{\pm} = - \Omega_{\pm}$ and $\Lambda_{0} = - \Omega_{0}$. Consequently, the equalities $\Omega_{\pm} = - \Omega_{\mp}^{\ast}$ 
and $\Omega_{0} = - \Omega_{0}^{\ast}$ are satisfied only for $\Omega_{+} = \xi$, $\Omega_{-} = - \xi^{\ast}$, and $\Omega_{0} = \nc \om$, 
with $\xi \in \mathbb{C}$ and $\om \in \mathbb{R}$. This same particular subset of arbitrary c-number parameters can also be obtained 
from the analysis of 
\bd
\lbk {\bf T}(\Omega_{\pm},\Omega_{0}) \rbk^{\dag} {\bf T}(\Omega_{\pm},\Omega_{0}) = {\bf 1} , 
\ed
which implies that ${\bf T}(\xi,\om)$ gives a family of unitary transformations for a general class of representations associated with 
the $\mathfrak{su}(1,1)$ Lie algebra. $\blacksquare$
\end{proof}

To illustrate our results, let us consider as a first example the single-mode bosonic representation of the Heisenberg-Weyl algebra
where the generators ${\bf K}_{\pm}$ and ${\bf K}_{0}$ are expressed as ${\bf K}_{+} = (1/2) {\bf a}^{\dag 2}$, ${\bf K}_{-} = (1/2)
{\bf a}^{2}$, and ${\bf K}_{0} = (1/2) ( {\bf a}^{\dag} {\bf a} + 1/2 )$, with ${\bf a}$ and ${\bf a}^{\dag}$ being, respectively, the
boson annihilation and creation operators satisfying the well-known commutation relation $[ {\bf a},{\bf a}^{\dag} ] = {\bf 1}$. In this
case, the unitary operator ${\bf T}(\xi,\om)$ assumes the form
\be
\lb{e8}
{\bf T}_{1}(\xi,\om) = e^{\half [ \xi {\bf a}^{\dag 2} + \nc \om \lpar {\bf a}^{\dag} {\bf a} + 1/2 \rpar - \xi^{\ast} {\bf a}^{2} ] } ,
\ee
which coincides with the squeezing operator
\bd
{\bf S}(\xi) \coloneq e^{\half ( \xi {\bf a}^{\dag 2} - \xi^{\ast} {\bf a}^{2} )} \quad \mbox{when} \quad \om = 0 . 
\ed
It is worth mentioning that, in particular, equation (\ref{e8}) can be promptly used to derive the quantum analogous of the Fresnel 
transform in classical optics \cite{Hong-Yi}.

Another interesting example encompasses the two-mode bosonic representation in which the generators are specifically given by
${\bf K}_{+} = {\bf a}^{\dag} {\bf b}^{\dag}$, ${\bf K}_{-} = {\bf a} {\bf b}$, and ${\bf K}_{0} = (1/2) ({\bf a}^{\dag} {\bf a} + 
{\bf b}^{\dag} {\bf b} + 1)$, where $[ {\bf a},{\bf a}^{\dag} ] = [ {\bf b},{\bf b}^{\dag} ] = {\bf 1}$. In this context, the unitary
operator
\be
\lb{e9}
{\bf T}_{2}(\xi,\om) = e^{\xi {\bf a}^{\dag} {\bf b}^{\dag} + \nc (\om/2) ({\bf a}^{\dag} {\bf a} + {\bf b}^{\dag} {\bf b} + 1) -
\xi^{\ast} {\bf a} {\bf b}}
\ee
recovers the two-mode squeezing operator 
\bd
{\bf S}_{2}(\xi) = \exp ( \xi {\bf a}^{\dag} {\bf b}^{\dag} - \xi^{\ast} {\bf a} {\bf b} ) \quad \mbox{for} \quad \om = 0 . 
\ed
Moreover, the action of ${\bf T}_{2}(\xi,\om)$ in the annihilation operators for each mode of the electromagnetic field implies in the 
following results:
\brr
\lb{e10}
{\bf T}_{2}^{\dag}(\xi,\om) {\bf a} {\bf T}_{2}(\xi,\om) &=& \lbk \cosh (\phi) + \nc (\om / 2 \phi) \sinh (\phi) \rbk {\bf a} \nn \\
& & + \, (\xi / \phi) \sinh (\phi) {\bf b}^{\dag} , \\
\lb{e11}
{\bf T}_{2}^{\dag}(\xi,\om) {\bf b} {\bf T}_{2}(\xi,\om) &=& \lbk \cosh (\phi) + \nc (\om / 2 \phi) \sinh (\phi) \rbk {\bf b} \nn \\
& & + \, (\xi / \phi) \sinh (\phi) {\bf a}^{\dag} ,
\err
being $\phi = [ | \xi |^{2} - ( \om/2 )^{2} ]^{1/2}$. This set of unitary transformations for $\om = 0$ has its counterpart in the
quantum description of physical processes involving parametric amplification \cite{Yariv,Mollow}. Recently, Pielawa {\it et al}.
\cite{Vitali} have proposed a new method for generating two-mode squeezing in high-Q resonators using a beam of atoms (which acts as a
reservoir for the field) with random arrival times. In particular, the authors have used the unitary two-mode squeezing operator 
${\bf S}_{2}(\xi)$ to bring an effective Hamiltonian -- which describes resonant single-photon processes -- to the well-known
Jaynes-Cummings form, where the new bosonic operators are connected to the old ones by two-mode squeezing transformations. Another
interesting application is based on the analogy between phonons in an axially time-dependent ion trap and quantum fields in an
expanding/contracting universe, where the multimode squeezing operator represents the basic mechanism for cosmological particle
creation \cite{Ion}. 

It is worth noticing that there are some textbooks on quantum optics \cite{Books} which discuss particular cases (if one compares with
those exposed here) of one- and two-mode squeezing operators in different contexts in physics and their connections with nonclassical 
states of the electromagnetic field. However, the dynamical origin of the real parameter $\om$ present in equation (\ref{e9}), for 
example, has not been investigated up until now in literature, and this fact will be our object of study in the next section. For this 
intent, we obtain the exact solutions of the Heisenberg equations for a specific nonresonant system like the parametric amplifier, and 
show the link between time-evolution operator and ${\bf T}_{2}(\xi,\om)$. 

\section{Parametric amplifier}

The parametric amplifier model proposed by Louisell {\it et al}. \cite{Yariv} consists basically of two coupled modes of the
electromagnetic field, which play a symmetrical role in the amplification process \cite{Mollow}. Such dynamical elements are usually
described by the Hamiltonian ($\hbar = 1$)
\be
\lb{e12}
{\bf H}(t) = \om_{a} {\bf a}^{\dag} {\bf a} + \om_{b} {\bf b}^{\dag} {\bf b} + \kappa \, {\bf a} {\bf b} \, e^{\nc \eta t} + 
\kappa^{\ast} {\bf a}^{\dag} {\bf b}^{\dag} e^{- \nc \eta t} ,
\ee
where ${\bf a}$ $({\bf b})$ is the annihilation operator for the signal (idler) mode, $\eta$ provides the frequency of the pump field
(which has been assumed strong enough to be expressed in classical terms), and $\kappa$ represents a complex coupling constant being
proportional to the second-order susceptibility of the nonlinear medium and to the amplitude of the pump. Moreover, let us introduce
a small deviation $\delta$ in the usual definition of $\eta$ such that $\eta = \om_{a} + \om_{b} + \delta$ with $\delta/ ( \om_{a} +
\om_{b}) \ll 1$ (by hypothesis we are assuming that $\delta$ comes from a non-perfect match between the frequencies $\eta$ and
$\om_{a} + \om_{b}$). In this case, the solutions of the Heisenberg equations are given by
\brr
{\bf a}(t) &=& e^{- \nc (\om_{a} + \delta/2) t} \Bigl\{ \lbk \cosh (\varphi t) + \nc (\delta/ 2 \varphi) \sinh (\varphi t) \rbk 
{\bf a}(0) \nn \\
& & - \nc (\kappa^{\ast} / \varphi) \sinh (\varphi t) {\bf b}^{\dag}(0) \Bigr\} , \nn \\
{\bf b}(t) &=& e^{- \nc (\om_{b} + \delta/2) t} \Bigl\{ \lbk \cosh (\varphi t) + \nc (\delta/ 2 \varphi) \sinh (\varphi t) \rbk 
{\bf b}(0) \nn \\
& & - \nc (\kappa^{\ast} / \varphi) \sinh (\varphi t) {\bf a}^{\dag}(0) \Bigr\} , \nn
\err
plus their Hermitian conjugates for $\varphi = [ | \kappa |^{2} - (\delta/2)^{2} ]^{1/2}$ fixed. These solutions lead us to verify that
$[ {\bf a}(t),{\bf a}^{\dag}(t) ] = [ {\bf b}(t),{\bf b}^{\dag}(t) ] = {\bf 1}$, which implies in the unitariety of the time-evolution 
operator ${\bf U}(t)$. This fact leads us to show that ${\bf n}_{a}(t) - {\bf n}_{b}(t) = {\bf n}_{a}(0) - {\bf n}_{b}(0)$ (conservation
law) for ${\bf n}_{c}(t) \coloneq {\bf c}^{\dag}(t) {\bf c}(t)$ is easily seen to hold; therefore, the intensity correlation function 
$\lgg {\bf n}_{a}(t) {\bf n}_{b}(t) \rg$ can be written as $\lgg {\bf n}_{a}(t) {\bf n}_{b}(t) \rg = \lgg {\bf n}_{a}^{2}(t) \rg + \lgg 
{\bf n}_{a}(t) [ {\bf n}_{b}(0) - {\bf n}_{a}(0) ] \rg$. For instance, when the initial state coincides with the number states 
$\{ | n_{a},n_{b} \rg \}_{n_{a},n_{b} \in \mathbb{N}}$, the second term on the rhs of this equation is reduced to $(n_{b}-n_{a}) \lgg 
n_{a},n_{b} | {\bf n}_{a}(t) | n_{a},n_{b} \rg$ for $n_{a} \neq n_{b}$. So, if one considers $n_{a} = n_{b}$, there is no contribution 
from this term and consequently, $\lgg {\bf n}_{a}(t) {\bf n}_{b}(t) \rg$ corresponds to the maximum violation of the Cauchy-Schwarz 
inequality $\lgg {\bf a}^{\dag} {\bf a} {\bf b}^{\dag} {\bf b} \rg \leq \lgg {\bf a}^{\dag 2} {\bf a}^{2} \rg$ for the parametric 
amplifier \cite{Books}.

Next, we show how the time-evolution operator ${\bf U}(t)$ can be connected with the unitary operator ${\bf T}_{2}(\xi,\om)$. To
carry out this task, let us initially mention that the time-dependent global phase factors present in the solutions of the Heisenberg
equations are obtained through the action of the rotation operator
\be
\lb{e13}
{\bf R}(\om_{a},\om_{b},\delta;t) = e^{- \nc t \lbk \lpar \om_{a} + \delta/2 \rpar {\bf a}^{\dag} {\bf a} + \lpar \om_{b} + \delta/2 
\rpar {\bf b}^{\dag} {\bf b} \rbk} 
\ee
on the annihilation operators ${\bf a}(0)$ and ${\bf b}(0)$. The next step then consists in noticing that for $\phi = \varphi t$, 
$\om = \delta t$, and $\xi = - \nc \kappa^{\ast} t$, the operator
\be
\lb{e14}
{\bf T}_{2}(\kappa,\delta;t) = e^{- \nc t \lbk \kappa^{\ast} {\bf a}^{\dag} {\bf b}^{\dag} - (\delta/2) ({\bf a}^{\dag} {\bf a} + 
{\bf b}^{\dag} {\bf b} + 1) + \kappa \, {\bf a} {\bf b} \rbk} 
\ee
is responsible for generating the terms between braces in the solutions ${\bf a}(t)$ and ${\bf b}(t)$ -- see equations (\ref{e10}) and 
(\ref{e11}). After these considerations, it is easy to show that
\be
\lb{e15}
{\bf U}(t) \coloneq {\bf R}(\om_{a},\om_{b},\delta;t) {\bf T}_{2}(\kappa,\delta;t)
\ee
provides the solutions obtained above by means of the mathematical operation sketched in the identities ${\bf a}(t) = {\bf U}^{\dag}(t)
{\bf a}(0) {\bf U}(t)$ and ${\bf b}(t) = {\bf U}^{\dag}(t) {\bf b}(0) {\bf U}(t)$. Note that the particular nonresonant system here
studied represents a first dynamical application of the results obtained in Section II for the two-mode bosonic representation, where
the connection between the time-evolution operator (\ref{e15}) and the unitary operator (\ref{e9}) is promptly established. It is worth
emphasizing that different physical systems can also be used for explaining the dynamical origin of the real parameter $\om$ present in
${\bf T}_{2}(\xi,\om)$ (\eg, see Ref. \cite{Vitali}), but this fact still deserves be carefully investigated. 

\section{An initial study on entanglement via Wigner function}

Let us initially consider a specific subset of bipartite physical systems described by continuous variables such that the parametric
amplifier model here studied (or the simple model for parametric frequency conversion proposed by Louisell {\it et al}. \cite{Yariv}, and
subsequently studied in detail by Tucker and Walls \cite{Walls}) constitutes a particular element. Furthermore, let us also assume that
$\ro(t) = {\bf U}(t) \ro(0) {\bf U}^{\dagger}(t)$ describes the dynamics of any bipartite system belonging to this subset whose unitary
time-evolution operator ${\bf U}(t)$ is related to the Hamiltonian operator ${\bf H}(t)$; by hypothesis, the initial density operator
$\ro(0)$ represents the system prepared at the initial instant $t=0$ in any disentangled (entangled) state. The symmetric characteristic
function for this class of bipartite physical systems is given by $\mathscr{C}(\mathds{G};t) \coloneq \mathrm{Tr} [ {\bf D}(\mathds{G})
\ro(t) ]$, where ${\bf D}(\mathds{G}) \coloneq \exp ( - \mathds{G}^{\dagger} \mathds{EO} )$ defines a displacement operator written in
terms of the matrices $\mathds{G}^{\dagger} = \left( \begin{array}{cccc} \xi_{a}^{\ast} & \xi_{a} & \xi_{b}^{\ast} & \xi_{b} \end{array}
\right)$, $\mathds{E} = \mathds{I} \otimes \mathds{S}$ with $\mathds{I} = \mathrm{diag}(1,1)$ and $\mathds{S} = \mathrm{diag}(1,-1)$
being the respective $2 \times 2$ identity and symplectic matrices, and $\mathds{O}^{\dagger} = \left( \begin{array}{cccc} 
{\bf a}^{\dagger} & {\bf a} & {\bf b}^{\dagger} & {\bf b} \end{array} \right)$. Note that due to the cyclic property of the trace
operation, the symmetric characteristic function can also be evaluated through the mathematical statement $\mathrm{Tr} [ 
{\bf U}^{\dagger}(t) {\bf D}(\mathds{G}) {\bf U}(t) \ro(0) ]$, where the time evolution of the displacement operator plays an important
role.

Next, let us suppose that the action of the unitary time-evolution operator on the matrix $\mathds{O}$ transforms the annihilation
(creation) operators ${\bf a}$ $({\bf a}^{\dagger})$ and ${\bf b}$ $({\bf b}^{\dagger})$ following the general rule $\mathds{O}_{\hei}
(t) = {\bf U}^{\dagger}(t) \mathds{O} {\bf U}(t) = \mathds{T}(t) \mathds{O}$, where
\brr
\mathds{T}(t) = \left[ \begin{array}{cccc} \mu_{a}(t) & \nu_{a}(t)        & \chi_{a}(t)        & \eta_{a}(t) \\
                                    \nu_{a}^{\ast}(t) & \mu_{a}^{\ast}(t) & \eta_{a}^{\ast}(t) & \chi_{a}^{\ast}(t) \\
                                           \mu_{b}(t) & \nu_{b}(t)        & \chi_{b}(t)        & \eta_{b}(t) \\
                                    \nu_{b}^{\ast}(t) & \mu_{b}^{\ast}(t) & \eta_{b}^{\ast}(t) & \chi_{b}^{\ast}(t)
\end{array} \right] \nn
\err
represents a $4 \times 4$ matrix whose elements are c-number functions determined from the Heisenberg equations for $\mathds{O}_{\hei}
(t)$ with specific initial conditions that preserve the quantum mechanics (the subscript $H$ indicates that operators are in the
Heisenberg picture). For instance, when $t=0$ the c-number functions should necessarily imply in the mathematical identity
$\mathds{T}(0) = \mathrm{diag}(1,1,1,1)$; while for $t \geq 0$ the commutation relations
\brr
\lbk {\bf a}_{\hei}(t),{\bf a}_{\hei}^{\dagger}(t) \rbk &=& {\bf 1} , \quad \lbk {\bf b}_{\hei}(t),{\bf b}_{\hei}^{\dagger}(t) 
\rbk = {\bf 1} , \nn \\ 
\lbk {\bf a}_{\hei}(t),{\bf b}_{\hei}(t) \rbk &=& 0 , \quad \lbk {\bf a}_{\hei}(t),{\bf b}_{\hei}^{\dagger}(t) \rbk = 0 , \nn
\err
provide extra relations for the c-number functions which permit us to solve completely the Heisenberg equations, namely
\brr
| \mu_{a}(t) |^{2} - | \nu_{a}(t) |^{2} + | \chi_{a}(t) |^{2} - | \eta_{a}(t) |^{2} &=& 1 , \nn \\
| \mu_{b}(t) |^{2} - | \nu_{b}(t) |^{2} + | \chi_{b}(t) |^{2} - | \eta_{b}(t) |^{2} &=& 1 , \nn \\
\mu_{a}(t) \nu_{b}(t) - \nu_{a}(t) \mu_{b}(t) + \chi_{a}(t) \eta_{b}(t) - \eta_{a}(t) \chi_{b}(t) &=& 0 , \nn \\
\mu_{a}(t) \mu_{b}^{\ast}(t) - \nu_{a}(t) \nu_{b}^{\ast}(t) + \chi_{a}(t) \chi_{b}^{\ast}(t) - \eta_{a}(t) \eta_{b}^{\ast}(t) &=& 0 . \nn
\err

The first immediate consequence of these results is that ${\bf U}^{\dagger}(t) {\bf D}(\mathds{G}) {\bf U}(t)$ will produce a new 
displacement operator ${\bf D}(\mathds{Y}) = \exp (- \mathds{Y}^{\dagger} \mathds{EO})$ with $\mathds{Y}^{\dagger} = \left( 
\begin{array}{cccc} \bet_{a}^{\ast} & \bet_{a} & \bet_{b}^{\ast} & \bet_{b} \end{array} \right)$, where the new elements are connected 
with the old ones through the equalities
\brr
\bet_{a} &=& \mu_{a}^{\ast}(t) \xi_{a} - \nu_{a}(t) \xi_{a}^{\ast} + \mu_{b}^{\ast}(t) \xi_{b} - \nu_{b}(t) \xi_{b}^{\ast} , \nn \\
\bet_{b} &=& \chi_{a}^{\ast}(t) \xi_{a} - \eta_{a}(t) \xi_{a}^{\ast} + \chi_{b}^{\ast}(t) \xi_{b} - \eta_{b}(t) \xi_{b}^{\ast} , \nn
\err
plus their respective complex conjugates. For convenience in our calculations, $\bet_{a(b)}$ means a short notation for $\bet_{a(b)} =
\bet_{a(b)}(\xi_{a},\xi_{b};t)$. It is worth noticing that $\mathscr{C}(\mathds{G};t)$ may now be written as $\mathscr{C}(\mathds{Y};0)=
\mathrm{Tr} [ {\bf D}(\mathds{Y}) \ro(0) ]$ (\ie, the symmetric characteristic function $\mathscr{C}(\mathds{G};t)$ is thus specified
in terms of the form it takes at $t=0$, which corroborates the results obtained by Mollow and Glauber \cite{Mollow} for the parametric
amplifier model), and this fact will bring some insights into the study of entanglement for bipartite systems via Wigner function.

The Wigner function for this particular subset of bipartite physical systems may then be defined as the four-dimensional Fourier 
transform of the symmetric characteristic function $\mathscr{C}(\mathds{G};t)$, that is,
\be
\lb{e16}
\mathscr{W}(\mathds{X};t) = \int \frac{d^{2} \xi_{a} d^{2} \xi_{b}}{\pi^{2}} \exp \lpar \mathds{G}^{\dagger} \mathds{EX} \rpar
\mathscr{C}(\mathds{G};t) ,
\ee
with $\mathds{X}^{\dagger} \coloneq \left( \begin{array}{cccc} \alf_{a}^{\ast} & \alf_{a} & \alf_{b}^{\ast} & \alf_{b} \end{array} \right)$. 
Note that the variables of integration $\xi_{a}$ and $\xi_{b}$ can be changed to $\bet_{a}$ and $\bet_{b}$ in this equation, once the 
Jacobian matrix of the transformation has a determinant whose absolute value is equal to one -- this fact implies in the identity 
$d^{2} \xi_{a} d^{2} \xi_{b} = d^{2} \bet_{a} d^{2} \bet_{b}$. Hence, the integration can now be conveniently carried out through the 
mathematical relation
\brr
\lb{e17}
\mathscr{W}(\mathds{X};t) &=& \int \frac{d^{2} \bet_{a} d^{2} \bet_{b}}{\pi^{2}} \exp \lpar \mathds{Y}^{\dagger} \mathds{EZ} \rpar
\mathscr{C}(\mathds{Y};0) \nn \\
&=& \mathscr{W}(\mathds{Z};0) ,
\err
where $\mathds{Z}^{\dagger} \coloneq \left( \begin{array}{cccc} \gam_{a}^{\ast} & \gam_{a} & \gam_{b}^{\ast} & \gam_{b} \end{array}
\right)$ defines a new matrix with elements given by\footnote{Note that the extra relations obtained from the commutation relations for 
the time-dependent c-number functions lead us to determine the inverse matrix $\mathds{T}^{-1}(t)$ and consequently, to establish the 
identities $\mathds{Y} = \mathds{T}^{-1}(t) \mathds{G}$ and $\mathds{Z} = \mathds{T}^{-1}(t) \mathds{X}$.}
\brr
\gam_{a} &=& \mu_{a}^{\ast}(t) \alf_{a} - \nu_{a}(t) \alf_{a}^{\ast} + \mu_{b}^{\ast}(t) \alf_{b} - \nu_{b}(t) \alf_{b}^{\ast}, \nn \\
\gam_{b} &=& \chi_{a}^{\ast}(t) \alf_{a} - \eta_{a}(t) \alf_{a}^{\ast} + \chi_{b}^{\ast}(t) \alf_{b} - \eta_{b}(t) \alf_{b}^{\ast}, \nn
\err
and their respective complex conjugates. Thus, equation (\ref{e17}) asserts that $\mathscr{W}(\mathds{X};t)$ can also be expressed in
terms of the form it takes at the initial time $t=0$, since the new variables $\gam_{a}(\alf_{a},\alf_{b};t)$ and $\gam_{b}(\alf_{a},
\alf_{b}; t)$ carry the information of the {\em dynamical entanglement} between the variables $\alf_{a}$ and $\alf_{b}$. Furthermore, this
result permits us to show that for a given bipartite system initially prepared in any entangled state, it will remain entangled for all 
$t \geq 0$; otherwise, if the density operator is described at $t=0$ as $\ro(0) = \ro_{a}(0) \otimes \ro_{b}(0)$ (disentangled state), 
the appearance of entanglement in the Wigner function will depend exclusively on the dynamics provided by the Hamiltonian operator 
${\bf H}(t)$\footnote{According to Mollow and Glauber \cite{Mollow}: `The fact that the Wigner function has this property is a
consequence of the form taken by the Hamiltonian (\ref{e12}). It may be shown that whenever the Hamiltonian of a system of oscillators
is given by a quadratic form in the creation and annihilation operators, the Wigner function is constant along classical trajectories.
This property does not extend to systems with arbitrary Hamiltonians, as it does in the case of the classical phase-space distribution.'
Beyond these fundamental features, it is worth mentioning that recent studies on the characterization and quantification of entanglement
for symmetric and asymmetric bipartite Gaussian states have contributed considerably to our comprehension of this important nonclassical
effect in quantum optics and quantum information theory \cite{Englert,Dodonov}. In this sense, we believe that the formalism here
presented for the Wigner function can help to extend such results in order to include some important dynamical effects that allow us to
better understand certain peculiarities on the entanglement process for a specific subset of bipartite physical systems described by
continuous variables \cite{Braunstein}.} -- here associated with the time-dependent matrix $\mathds{Z}$. Following, let us apply the
results obtained until now to the parametric amplifier model, where different initial states of the electromagnetic field will be
considered.

Now we focus our efforts in evaluating $\mathscr{W}(\mathds{Z};0)$ for a well-known family of two-mode electromagnetic fields where, in
particular, both modes are initially prepared in the coherent states, number states, and thermal states, respectively. For instance,
the symmetric characteristic functions in these situations are given by
\brr
\mathscr{C}_{\mathrm{coh}}(\mathds{Y};0) &=& e^{- \half \lpar | \bet_{a} |^{2} + | \bet_{b} |^{2} \rpar + 2 \nc \lbk 
\mathrm{Im} ( \bet_{a} \zeta_{a}^{\ast} ) + \mathrm{Im} ( \bet_{b} \zeta_{b}^{\ast} ) \rbk} , \nn \\
\mathscr{C}_{\mathrm{n}}(\mathds{Y};0) &=& e^{- \half \lpar | \bet_{a} |^{2} + | \bet_{b} |^{2} \rpar} L_{n_{a}}
( | \bet_{a} |^{2} ) L_{n_{b}} ( | \bet_{b} |^{2} ) , \nn \\
\mathscr{C}_{\mathrm{th}}(\mathds{Y};0) &=& e^{- \half \lbk ( 1 + 2 \bar{n}_{a} ) | \bet_{a} |^{2} + (1 + 2 \bar{n}_{b}) 
| \bet_{b} |^{2} \rbk} , \nn
\err
while their respective Wigner functions can be expressed as
\brr
\lb{e18}
\!\!\!\!\!\!\!\!\!\! \mathscr{W}_{\mathrm{coh}}(\mathds{Z};0) &=& 4 \, e^{- 2 \lpar | \gam_{a} - \zeta_{a} |^{2} + | \gam_{b} - 
\zeta_{b} |^{2} \rpar} , \\
\lb{e19}
\!\!\!\!\!\!\!\!\!\! \mathscr{W}_{\mathrm{n}}(\mathds{Z};0) &=& 4 (-1)^{n_{a}+n_{b}} e^{- 2 \lpar | \gam_{a} |^{2} + | \gam_{b} |^{2} 
\rpar} L_{n_{a}} (4 | \gam_{a} |^{2} ) \nn \\
& & \times \, L_{n_{b}} (4 | \gam_{b} |^{2} ) , \\
\lb{e20}
\!\!\!\!\!\!\!\!\!\! \mathscr{W}_{\mathrm{th}}(\mathds{Z};0) &=& 4 \lbk (1 + 2 \bar{n}_{a}) (1 + 2 \bar{n}_{b}) \rbk^{-1} \nn \\ 
& & \times \, e^{- 2 \lbk (1 + 2 \bar{n}_{a})^{-1} | \gam_{a} |^{2} + (1 + 2 \bar{n}_{b})^{-1} | \gam_{b} |^{2} \rbk} ,
\err
with $L_{n}(z)$ denoting a Laguerre polynomial and $\bar{n}_{a(b)}$ being the mean number of photons for a chaotic light 
field. Note that if we consider the solutions obtained from the Heisenberg equations for the parametric amplifier model (see previous
section), it is easy to show that $\mu_{a}(t)$, $\eta_{a}(t)$, $\nu_{b}(t)$, and $\chi_{b}(t)$ (once the further time-dependent
coefficients do no exist) determine completely the c-numbers $\bet_{a(b)}(\xi_{a},\xi_{b};t)$ and $\gam_{a(b)}(\alf_{a},\alf_{b};t)$.
Indeed, this last step leads us to characterize precisely $\mathscr{C}(\mathds{Y};0)$ and $\mathscr{W}(\mathds{Z};0)$. Moreover, when
$\delta = 0$, $\kappa \in \mathbb{R}_{+}$, and $(\om_{a} + \om_{b})t = 3 \pi/2$, Eq. (\ref{e18}) coincides with $\mathscr{W}_{\epr}
(\gam_{a},\gam_{b})$ for $\zeta_{a(b)} = 0$ (two-mode vacuum state), this function being that used by Braunstein and Kimble
\cite{Kimble} in the theoretical description of teleportation involving continuous quantum variables.\footnote{It is worth noticing that
the cases here studied represent a specific set of disentangled bipartite states whose characteristic and Wigner functions are written
as a product of two functions associated with each mode separately, \ie, since $\ro(0) = \ro_{a}(0) \otimes \ro_{b}(0)$ we promptly
obtain the relations $\mathscr{C}(\mathds{Y};0) = \mathscr{C}_{a}(\bet_{a};0) \mathscr{C}_{b}(\bet_{b};0)$ and $\mathscr{W}
(\mathds{Z};0) = \mathscr{W}_{a}(\gam_{a};0) \mathscr{W}_{b}(\gam_{b};0)$. In this situation, the dynamical entanglement originated from
the Hamiltonian ${\bf H}(t)$ allows us to correlate the complex variables $\alf_{a}$ and $\alf_{b}$ (or $\xi_{a}$ and $\xi_{b}$) present
in $\gam_{a(b)}$ $(\bet_{a(b)})$.}

Finally, let us say some few words about the results obtained in this section. It is important to emphasize that equation (\ref{e17}) 
permits us to describe qualitatively the entanglement of a specific subset of bipartite systems, where now we can promptly identify 
two distinct origins of this quantum effect: the first associated with the entanglement in the initial conditions (here labelled by 
kinematical entanglement), while the second is responsible for the dynamical entanglement via Hamiltonian operator. In addition, the 
normally ordered moments
\brr
\lb{e21}
& & \lgg {\bf a}^{\dagger p}(t) {\bf a}^{q}(t) {\bf b}^{\dagger r}(t) {\bf b}^{s}(t) \rg = \bgam_{\xi_{{\rm a}},\xi_{{\rm a}}^{\ast},
\xi_{{\rm b}},\xi_{{\rm b}}^{\ast}}^{(p,q,r,s)} e^{\half \lpar | \xi_{{\rm a}} |^{2} + | \xi_{{\rm b}} |^{2} \rpar} \nn \\
& & \qquad \qquad \times \, \mathscr{C}(\xi_{a},\xi_{a}^{\ast},\xi_{b},\xi_{b}^{\ast};t) \biggr|_{\xi_{{\rm a}},\xi_{{\rm a}}^{\ast},
\xi_{{\rm b}},\xi_{{\rm b}}^{\ast} = 0} ,
\err
with
\bd
\bgam_{\xi_{{\rm a}},\xi_{{\rm a}}^{\ast},\xi_{{\rm b}},\xi_{{\rm b}}^{\ast}}^{(p,q,r,s)} \coloneq (-1)^{q+s} \frac{\upartial^{p+q+r+s}}
{\upartial \xi_{{\rm a}}^{p} \upartial \xi_{{\rm a}}^{\ast q} \upartial \xi_{{\rm b}}^{r} \upartial \xi_{{\rm b}}^{\ast s}}
\ed
and $\{ p,q,r,s \} \in \mathbb{N}$, can also be used to investigate some recent proposals of inseparability criteria for continuous
bipartite quantum states \cite{Dodonov,JCM}. In the next section, we will establish a reasonable measure of entanglement which is based
on the results obtained by Pi\c{a}tek and Leo\'{n}ski \cite{Piatek} for the intermode correlations in phase space. 

\section{Entropy functionals for continuous phase-space representations}

In order to establish a quantitative estimate of entanglement for bipartite systems, let us introduce some basic definitions of entropy 
functionals in continuous phase-space representations. The first definition is based on the joint entropy \cite{Piatek}
\be
\lb{e22}
\mathrm{E}[\mathscr{H};t] \coloneq - \int \frac{d^{2} \alf_{a} d^{2} \alf_{b}}{\pi^{2}} \, \mathscr{H}(\alf_{a},\alf_{b};t) \ln \lbk
\mathscr{H}(\alf_{a},\alf_{b};t) \rbk ,
\ee
where $\mathscr{H}(\alf_{a},\alf_{b};t) \coloneq \lgg \alf_{a},\alf_{b} | \ro(t) | \alf_{a},\alf_{b} \rg$ denotes the Husimi function
in the continuous coherent-state representations for a bipartite system described by the density operator $\ro(t)$. Note that
$\mathrm{E}[\mathscr{H};t]$ presents certain properties inherent to its definition which deserve be mentioned: (i) the probability
distribution function $\mathscr{H}(\alf_{a},\alf_{b};t)$ is strictly positive and limited to the interval $[0,1]$; consequently, the
joint entropy (\ref{e22}) characterizes a well-defined function whose behaviour does not present any mathematical inconsistencies.
Moreover, (ii) this definition essentially measures the functional correlation between the continuous variables used to describe each
part of the joint system. Hence, $\mathrm{E}[\mathscr{H};t]$ can be considered as a natural extension of that definition employed by
Wehrl \cite{Wehrl} for information entropy.\footnote{Recently, Mintert and \.{Z}yczkowski \cite{Mintert} evaluated explicitly the Wehrl
and generalized R\'{e}nyi-Wehrl entropy functionals for any pure states describing $N \times N$ bipartite quantum systems. For this
intent, they properly defined the Husimi function for the $\mathfrak{su}(N) \times \mathfrak{su}(N)$ coherent-state representations,
and showed that: (i) the Wehrl entropy functional is minimal iff the pure states above mentioned are separable, (ii) the excess of this
quantity is equal to the subentropy of the mixed states obtained by the partial trace of the bipartite pure states; and finally, (iii)
these functionals can be considered as alternative measures of entanglement. Here, our intention is to establish an additional measure
of entanglement (limited to the interval $[0,1]$) for Wehrl's entropy functionals described by bipartite Husimi functions which admit
$\mathfrak{su}(1,1) \times \mathfrak{su}(1,1)$ coherent-state representations.}

The second definition consists of functionals related to the partial entropies 
\be
\lb{e23}
\mathrm{E}[\mathscr{H}^{(\inda)};t] \coloneq - \int \frac{d^{2} \alf_{a}}{\pi} \, \mathscr{H}^{(\inda)}(\alf_{a};t) \ln \lbk
\mathscr{H}^{(\inda)}(\alf_{a};t) \rbk
\ee
and
\be
\lb{e24}
\mathrm{E}[\mathscr{H}^{(\indb)};t] \coloneq - \int \frac{d^{2} \alf_{b}}{\pi} \, \mathscr{H}^{(\indb)}(\alf_{b};t) \ln \lbk
\mathscr{H}^{(\indb)}(\alf_{b};t) \rbk ,
\ee
which depend basically on the marginal Husimi functions
\brr
\mathscr{H}^{(\inda)}(\alf_{a};t) &=& \int \frac{d^{2} \alf_{b}}{\pi} \mathscr{H}(\alf_{a},\alf_{b};t) , \nn \\
\mathscr{H}^{(\indb)}(\alf_{b};t) &=& \int \frac{d^{2} \alf_{a}}{\pi} \mathscr{H}(\alf_{a},\alf_{b};t) . \nn
\err
Such marginal Husimi functions carry information of the entanglement between the subsystems `A' and `B', since the partial trace
over the continuous variables $\alf_{a(b)}$ of a particular subsystem `A' (`B') allows the introduction, via time-evolution operator
${\bf U}(t)$ and/or initial density operator $\ro(0)$, of important correlations in the bipartite states. 

Let us derive now some mathematical relations among these entropy functionals from the Araki-Lieb inequality \cite{Araki}, \ie,
\brr
\lb{e25}
\left| \mathrm{E}[\mathscr{H}^{(\inda)};t] - \mathrm{E}[\mathscr{H}^{(\indb)};t] \right| &\leq& \mathrm{E}[\mathscr{H};t] \leq 
\mathrm{E}[\mathscr{H}^{(\inda)};t] \nn \\
&+& \mathrm{E}[\mathscr{H}^{(\indb)};t] .
\err
For instance, the rhs of this inequality corresponds to the subadditivity property for the Wehrl's entropy functionals, while the equal 
sign reflects the complete disentanglement between the subsystems `A' and `B' of the joint system. In addition, the conditional entropies 
\cite{Piatek,Wehrl}
\be
\lb{e26}
\mathrm{E}[\mathscr{H}/\mathscr{H}^{(\inda)};t] = \mathrm{E}[\mathscr{H};t] - \mathrm{E}[\mathscr{H}^{(\inda)};t]
\ee
and
\be
\lb{e27}
\mathrm{E}[\mathscr{H}/\mathscr{H}^{(\indb)};t] = \mathrm{E}[\mathscr{H};t] - \mathrm{E}[\mathscr{H}^{(\indb)};t]
\ee
lead us not only to establish the balance equation
\be
\lb{e28}
\mathrm{E}[\mathscr{H}/\mathscr{H}^{(\inda)};t] + \mathrm{E}[\mathscr{H}^{(\inda)};t] = \mathrm{E}[\mathscr{H}/
\mathscr{H}^{(\indb)};t] + \mathrm{E}[\mathscr{H}^{(\indb)};t] ,
\ee
but also to determine the further inequalities 
\bd
\mathrm{E}[\mathscr{H}/\mathscr{H}^{(\inda)};t] \leq \mathrm{E}[\mathscr{H}^{(\indb)}; t] 
\ed
and 
\bd
\mathrm{E}[\mathscr{H}/\mathscr{H}^{(\indb)};t] \leq \mathrm{E}[\mathscr{H}^{(\inda)};t]
\ed
from the subadditivity property. It is worth noticing that the equal signs hold in both situations only when the c-numbers $\alf_{a}$
and $\alf_{b}$ are functionally uncorrelated, namely, the bipartite Husimi function $\mathscr{H}(\alf_{a},\alf_{b};t)$ factorizes in
the product of the marginal Husimi functions $\mathscr{H}^{(\inda)}(\alf_{a};t) \mathscr{H}^{(\indb)}(\alf_{b};t)$.

In what concerns the class of entropic functionals listed until now, it is convenient to define a new functional limited to the closed
interval $[0,1]$ for any $t \geq 0$, which permits us to avoid any ambiguity in the significance of the subadditivity property. Thus,
let us introduce the correlation functional
\be
\lb{e29}
\mathrm{C}[\mathscr{H};t] \coloneq 2 \lpar 1 - \frac{\mathrm{E}[\mathscr{H};t]}{\mathrm{E}[\mathscr{H}^{(\inda)};t] +
\mathrm{E}[\mathscr{H}^{(\indb)};t]} \rpar ,
\ee
which can be used to measure the functional correlation (that is, the `degree of entanglement') between the parts `A' and `B' of the
joint system, having as reference a factorizable bipartite Husimi function $\mathscr{H}(\alf_{a},\alf_{b};t)$. In this definition, the
inferior limit of the closed interval $[0,1]$ corresponds to the situation where the bipartite Husimi function factorizes, while its
superior limit reflects the opposite situation (\ie, when the bipartite states are maximally correlated); besides, the global factor
`2' is associated with the number of possible combinations between the constituent parts of the system under investigation. As a first 
practical application of these entropy functionals we will consider two different examples of initially uncoupled bipartite states for 
the parametric amplifier model.

\subsection{Coherent states}

In this first example, we consider the bipartite pure states $\ro(0) = | \zeta_{a},\zeta_{b} \rg \lgg \zeta_{a},\zeta_{b} |$ as being
the initial state at $t=0$ of the model governed by the Hamiltonian (\ref{e12}). Here, $| \zeta_{a},\zeta_{b} \rg = | \zeta_{a} \rg
\otimes | \zeta_{b} \rg$ where $| \zeta_{a(b)} \rg$ characterizes the coherent states related to the signal (idler) mode of the 
electromagnetic field. Thus, after some nontrivial algebra, the bipartite Husimi function assumes the closed-form
\brr
\lb{e30}
\mathscr{H}_{\mathrm{coh}}(\alf_{a},\alf_{b};t) &=& | A_{0} | e^{- \lpar | \gam_{a} - \zeta_{a} |^{2} + | \gam_{b} - \zeta_{b} |^{2}
\rpar} \nn \\
& & \times \, e^{2 \mathrm{Re} \lbk A_{+}^{\ast} ( \gam_{a} - \zeta_{a} )( \gam_{b} - \zeta_{b} ) \rbk}
\err
such that
\brr
\gam_{a(b)} &=& e^{\nc (\om_{a(b)} + \delta/2) t} \lbk \cosh (\varphi t) - \nc (\delta / 2 \varphi) \sinh (\varphi t) \rbk 
\alf_{a(b)} \nn \\
& & + \, \nc (\kappa^{\ast} / \varphi) e^{- \nc (\om_{b(a)} + \delta/2) t} \sinh (\varphi t) \, \alf_{b(a)}^{\ast} , \nn \\
A_{+} &=& \frac{\nc (\kappa^{\ast} / \varphi) \sinh (\varphi t)}{\cosh (\varphi t) + \nc (\delta / 2 \varphi) \sinh (\varphi t)}, \nn \\
A_{0} &=& \lbk \cosh (\varphi t) + \nc (\delta / 2 \varphi) \sinh (\varphi t) \rbk^{-2} . \nn
\err
Furthermore, the marginal Husimi functions
\be
\lb{e31}
\mathscr{H}_{\mathrm{coh}}^{(\inda)}(\alf_{a};t) = | A_{0} | e^{- | A_{0} | | \alf_{a} - \eps_{a} |^{2}}
\ee
and
\be 
\lb{e32}
\mathscr{H}_{\mathrm{coh}}^{(\indb)}(\alf_{b};t) = | A_{0} | e^{- | A_{0} | | \alf_{b} - \eps_{b} |^{2}} ,
\ee
with
\brr
\eps_{a(b)} &=& \Bigl\{ \lbk \cosh (\varphi t) + \nc (\delta / 2 \varphi) \sinh (\varphi t) \rbk \zeta_{a(b)} \nn \\
& & - \nc ( \kappa^{\ast} / \varphi) \sinh (\varphi t) \zeta_{b(a)}^{\ast} \Bigr\} e^{- \nc (\om_{a(b)} + \delta/2) t} , \nn
\err
permit us to verify how the dynamical entanglement introduces correlations between the continuous variables $\zeta_{a(b)}$ and 
$\zeta_{b(a)}^{\ast}$ through the label $\eps_{a(b)}$. Indeed, for $t=0$ it is possible to demonstrate that $\mathscr{H}_{\mathrm{coh}}
(\alf_{a},\alf_{b};0)$ factorizes in the product $\mathscr{H}_{\mathrm{coh}}^{(\inda)}(\alf_{a};0) \mathscr{H}_{\mathrm{coh}}^{(\indb)}
(\alf_{b};0)$. However, if one considers $t > 0$, the dynamics governed by the process of parametric amplification introduces
significant correlations between the signal and idler modes. Following, let us quantify these correlations with the help of Eq. (\ref{e29}).

The Gaussian structures present in the bipartite and marginal Husimi functions allow us to obtain closed form expressions for the
respective joint and partial entropies, namely,
\be
\lb{e33}
\mathrm{E}[\mathscr{H}_{\mathrm{coh}};t] = 2 + \ln \lbk 1 + ( | \kappa | / \varphi )^{2} \sinh^{2} ( \varphi t ) \rbk
\ee
and
\be
\lb{e34}
\mathrm{E}[\mathscr{H}_{\mathrm{coh}}^{(\inda, \indb)};t] = 1 + \ln \lbk 1 + ( | \kappa | / \varphi )^{2} \sinh^{2} ( \varphi t ) \rbk .
\ee
Consequently, the correlation functional can be promptly evaluated as follows:
\be
\lb{e35}
\mathrm{C}[\mathscr{H}_{\mathrm{coh}};t] = \frac{\ln \lbk 1 + ( | \kappa | / \varphi )^{2} \sinh^{2} ( \varphi t ) \rbk}{1 + \ln \lbk 
1 + ( | \kappa | / \varphi )^{2} \sinh^{2} ( \varphi t ) \rbk} .
\ee
Besides, if one considers the limit $\varphi t \gg 1$ for sufficiently long time, the correlation functional goes asymptotically to
$1$, this value being associated with the specific dynamics given by the Hamiltonian (\ref{e12}). On the other hand, for 
$\varphi t = 0$ we obtain $\mathrm{C}[\mathscr{H}_{\mathrm{coh}};0] = 0$, which corroborates the factorization process of the bipartite
Husimi function $\mathscr{H}_{\mathrm{coh}}(\alf_{a},\alf_{b};0)$.

\subsection{Thermal states}

Now, let us suppose that both signal and idler modes were initially prepared in the thermal states. In this case, the initial density
operator admits the expansion
\bd
\ro_{\mathrm{th}}(0) = \frac{1}{\bar{n}_{a} \bar{n}_{b}} \int \frac{d^{2} \zeta_{a} d^{2} \zeta_{b}}{\pi^{2}} e^{- \lpar \bar{n}_{a}^{-1} 
| \zeta_{a} |^{2} + \bar{n}_{b}^{-1} | \zeta_{b} |^{2} \rpar} {\bf P}(\zeta_{a},\zeta_{b}) ,
\ed
where $\bar{n}_{a(b)}$ represents the average occupation number of each mode, and ${\bf P}(\zeta_{a},\zeta_{b}) = | \zeta_{a},\zeta_{b} 
\rg \lgg \zeta_{a},\zeta_{b} |$ is the projector of the coherent states. This typical example of bipartite mixed states constitutes
an important starting point in the investigation process of the dynamical entanglement due to the parametric amplifier model. So, after
some calculations, the bipartite Husimi function can be written as
\brr
\lb{e36}
\mathscr{H}_{\mathrm{th}}(\alf_{a},\alf_{b};t) &=& \lbk \mathscr{G}_{\bar{n}_{a},\bar{n}_{b}}(1,1;t) \rbk^{-1} e^{- \lpar \ell_{b} 
| \alf_{a} |^{2} + \ell_{a} | \alf_{b} |^{2} \rpar} \nn \\
& & \times \, e^{2 \mathrm{Re} (\ell_{ab} \alf_{a} \alf_{b})}
\err
with
\brr
\ell_{a} &=& \frac{\mathscr{G}_{\bar{n}_{a},\bar{n}_{b}}(1,0;t)}{\mathscr{G}_{\bar{n}_{a},\bar{n}_{b}}(1,1;t)} , \qquad
\ell_{b} = \frac{\mathscr{G}_{\bar{n}_{a},\bar{n}_{b}}(0,1;t)}{\mathscr{G}_{\bar{n}_{a},\bar{n}_{b}}(1,1;t)} , \nn \\
\ell_{ab} &=& \nc ( \kappa / \varphi ) \sinh (\varphi t) \lbk \cosh (\varphi t) - \nc ( \delta / 2 \varphi ) \sinh (\varphi t) \rbk \nn \\
& & \times \lpar \bar{n}_{a} + \bar{n}_{b} + 1 \rpar \lbk \mathscr{G}_{\bar{n}_{a},\bar{n}_{b}}(1,1;t) \rbk^{-1} e^{\nc ( \om_{a} + 
\om_{b} + \delta ) t} , \nn
\err
and
\brr
\mathscr{G}_{\bar{n}_{a},\bar{n}_{b}}(x,y;t) &=& (\bar{n}_{a} x + 1)(\bar{n}_{b} y + 1) \nn \\
& & + \, (\bar{n}_{a} + \bar{n}_{b} + 1)( | \kappa | / \varphi )^{2} \sinh^{2} (\varphi t) ; \nn 
\err
while its joint entropy assumes the closed-form
\be
\lb{e37}
\mathrm{E}[\mathscr{H}_{\mathrm{th}};t] = 2 + \ln \lbk \mathscr{G}_{\bar{n}_{a},\bar{n}_{b}}(1,1;t) \rbk .
\ee
In addition, the marginal Husimi functions
\be
\lb{e38}
\mathscr{H}_{\mathrm{th}}^{(\inda)}(\alf_{a};t) = \lbk \mathscr{G}_{\bar{n}_{a},\bar{n}_{b}}(1,0;t) \rbk^{-1} 
e^{- \lbk \mathscr{G}_{\bar{n}_{a},\bar{n}_{b}}(1,0;t) \rbk^{-1} | \alf_{a} |^{2}}
\ee
and
\be
\lb{e39}
\mathscr{H}_{\mathrm{th}}^{(\indb)}(\alf_{b};t) = \lbk \mathscr{G}_{\bar{n}_{a},\bar{n}_{b}}(0,1;t) \rbk^{-1} 
e^{- \lbk \mathscr{G}_{\bar{n}_{a},\bar{n}_{b}}(0,1;t) \rbk^{-1} | \alf_{b} |^{2}}
\ee
permit also to determine explicitly the partial entropies
\brr
\lb{e40}
\mathrm{E}[\mathscr{H}_{\mathrm{th}}^{(\inda)};t] &=& 1 + \ln \lbk \mathscr{G}_{\bar{n}_{a},\bar{n}_{b}}(1,0;t) \rbk , \\
\lb{e41}
\mathrm{E}[\mathscr{H}_{\mathrm{th}}^{(\indb)};t] &=& 1 + \ln \lbk \mathscr{G}_{\bar{n}_{a},\bar{n}_{b}}(0,1;t) \rbk .
\err
In this moment becomes important noticing that the subadditivity property is not violated for any $t \geq 0$, the equality 
$\mathrm{E}[\mathscr{H}_{\mathrm{th}};0] = \mathrm{E}[\mathscr{H}_{\mathrm{th}}^{(\inda)};0] +
\mathrm{E}[\mathscr{H}_{\mathrm{th}}^{(\indb)};0]$ being consistent with the factorization of
$\mathscr{H}_{\mathrm{th}}(\alf_{a},\alf_{b};0)$ in the product $\mathscr{H}_{\mathrm{th}}^{(\inda)}(\alf_{a};0)
\mathscr{H}_{\mathrm{th}}^{(\indb)}(\alf_{b};0)$.

Following, let us investigate the correlation functional $\mathrm{C}[\mathscr{H}_{\mathrm{th}};t]$ for $t > 0$, since 
$\mathrm{C}[\mathscr{H}_{\mathrm{th}};0] = 0$ reflects the disentanglement between the bipartite mixed states. In fact, for $t \neq 0$
this functional assumes any values into the open interval $(0,1)$, the maximum value $1$ being reached only for sufficiently long
time. Hence, we can conclude that: (i) any initially uncoupled states become entangled in the process described by the Hamiltonian
(\ref{e12}), which confirms the results obtained by Dodonov {\it et al}. \cite{Dodonov} through the evaluation of the inverse negativity
coefficient (this measure also leads us to estimate quantitatively the `degree of entanglement' of a particular class of bipartite 
states such as those studied in this section); furthermore, (ii) the dynamics here characterized by the physical system under 
investigation produces a maximum estimate of entanglement for any initially disentangled states, this value (namely, $1$) being
considered as a specific quantum signature of the parametric amplifier model.

\section{Conclusions}

We have established a set of interesting formal results within the scope of quantum optics and quantum information theory that allows
us, among other things, 
\begin{itemize}
\item to define a new class of unitary squeezing transformations related to $\mathfrak{su}(1,1)$ Lie algebra which generalizes, for 
certain particular representations of its generators, the one- and two-mode squeezing operators \cite{Books}; 
\item to discuss, from the physical point of view, how the generalized two-mode squeezing operator can be generated through a slightly 
modified version of the parametric amplifier model \cite{Yariv}; 
\item to obtain a general integral representation for the bipartite Wigner function whose integrand is expressed as a product of two 
terms which are responsible for the dynamical and kinematical entanglement; and finally, 
\item to estimate quantitatively the `degree of entanglement' related to an ideal bipartite system (we are discarding the unvoidable 
coupling with the environment in this context) by means of a theoretical framework \cite{Piatek} which is based on the Wehrl's approach 
\cite{Wehrl} for the entropy functionals. 
\end{itemize}
In fact, equation (\ref{e29}) and its respective properties mentioned properly in the body of the text permit us to estimate, through an 
entropic approach, the entanglement effects for a wide class of electromagnetic field states, including {\em Gaussian} and 
{\em non-Gaussian} states. As a concluding remark, it is worth mentioning that these results have also potential applications in modern 
research on quantum teleportation \cite{Kimble}, quantum tomography \cite{Leonhardt,Marcelo}, and quantum computation \cite{Nielsen} 
(once continuous-variable entanglement can also be efficiently produced using squeezed light and linear optics \cite{Sanders}), as well 
as on the foundations of quantum mechanics through its extensions to the $\mathfrak{su}(2)$ Lie algebra \cite{Thomas}.

\section*{Acknowledgements}

This work has been supported by Conselho Nacional de Desenvolvimento Cient\'{\i}fico e Tecnol\'{o}gico (CNPq), Brazil.


\end{document}